\documentclass[letterpaper,11pt]{article}

\usepackage{amssymb}

\newcommand {\beq} {\begin{equation}}
\newcommand {\eeq} {\end{equation}}
\newcommand {\beqa} {\begin{eqnarray}}
\newcommand {\eeqa} {\end{eqnarray}}
\newcommand {\beqan} {\begin{eqnarray*}}
\newcommand {\eeqan} {\end{eqnarray*}}

\newcommand {\nn} {\nonumber}
\newcommand {\ph}[1]{\phantom{#1}}
\newcommand {\ie}{i.e.~}
\newcommand {\eg}{e.g.~}

\newcommand {\sss} {\scriptscriptstyle}

\newcommand{\al}{\ensuremath{\alpha}}
\newcommand{\be}{\ensuremath{\beta}}
\newcommand{\ga}{\ensuremath{\gamma}}

\newcommand{\de}{\ensuremath{\delta}}
\newcommand{\De}{\ensuremath{\Delta}}
\newcommand{\eps}{\ensuremath{\epsilon}}

\newcommand{\la}{\ensuremath{\lambda}}

\newcommand{\La}{\ensuremath{\Lambda}}
\newcommand{\si}{\ensuremath{\sigma}}

\newcommand{\Om}{\ensuremath{\Omega}}

\newcommand{\alh}{\ensuremath{\alpha}}
\newcommand{\beh}{\ensuremath{\beta}}
\newcommand{\gah}{\ensuremath{\gamma}}
\newcommand{\deh}{\ensuremath{\delta}}

\newcommand{\Psit}{\ensuremath{\widetilde{\Psi}}}
\newcommand{\PsiT}[1]{\ensuremath{\Psi^{\sss{T}}_{#1}}}
\newcommand{\PsitT}[1]{\ensuremath{\Psit^{\sss{T}}_{#1}}}

\newcommand{\Phih}{\ensuremath{\hat{\Phi}}}
\newcommand{\etat}{\ensuremath{\tilde{\eta}}}
\newcommand{\etaT}[1]{\ensuremath{{\eta^{{\sss T}}}^{#1}}}
\newcommand{\etaTe}[1]{\ensuremath{{\eta_1^{{\sss T}}}^{#1}}}
\newcommand{\etas}[1]{\ensuremath{{\eta^{*}}^{#1}}}
\newcommand{\etatT}[1]{\ensuremath{\tilde{\eta}^{{\sss T} #1}}}

\newcommand{\thetat}{\ensuremath{\tilde{\theta}}}
\newcommand{\thetatT}[1]{\ensuremath{\tilde{\theta}^{{\sss T} #1}}}

\newcommand{\Bf}[2]{B_{#1}^{\ph{#1}#2}}
\newcommand{\kde}[2]{\de_{#1}^{\ph{#1}#2}}

\newcommand{\mathHb}[1]{{\mathop{\kern0pt#1}\limits^{\,\sss
      \prime\prime}\vphantom{#1}}}

\newcommand{\phivev}{\ensuremath{\langle \phi \rangle}}

\newcommand{\Tr}[1]{\ensuremath{\mathrm{Tr}\left( #1 \right)}}

\newcommand{\com}[2]{\left[ #1 , #2 \right]}

\newcommand {\pa} {\partial}

\newcommand{\eqnlab}[1]{\label{eqn:#1}}

\newcommand{\eqnref}[1]{(\ref{eqn:#1})}

\newcommand{\Eqnref}[1]{Eq.~(\ref{eqn:#1})}

\newcommand{\Eqsref}[1]{Eqs.~(\ref{eqn:#1})}

\newcommand{\seclab}[1]{\label{sec:#1}}

\begin{document}

 \pagestyle{empty}
 \vskip-10pt
 \hfill {\tt hep-th/0306145}

\begin{center}

\vspace*{2cm}

\noindent
{\LARGE\textsf{\textbf{Free tensor multiplets and strings in
      spontaneously \\[5mm]
broken six-dimensional $(2, 0)$ theory}}}
\vskip 1truecm
\vskip 2truecm

{\large \textsf{\textbf{P\"ar Arvidsson\footnote{\tt par@fy.chalmers.se}, Erik
  Flink\footnote{\tt erik.flink@fy.chalmers.se} and M{\aa}ns
  Henningson\footnote{\tt mans@fy.chalmers.se}}}} \\ 
\vskip 1truecm
{\it Department of Theoretical  Physics\\ Chalmers University of
  Technology and G\"oteborg University\\ SE-412 96 G\"{o}teborg,
  Sweden}\\
\end{center}
\vskip 1cm
\noindent{\bf Abstract:}
We first review the representations of the six-dimensional $(2, 0)$
superalgebra on a free tensor multiplet and on a free string. We then
construct a supersymmetric Lagrangian describing a free tensor
multiplet. (It also includes a decoupled anti self-dual part of the
three-form field strength.) This field theory is then rewritten in
variables appropriate for analyzing a situation where the $R$-symmetry
is spontaneously broken by the vacuum expectation values of the scalar
moduli fields. Finally, we construct a supersymmetric and
$\kappa$-symmetric action for a free string.
\newpage
\pagestyle{plain}

\section{Introduction} 
In a previous publication~\cite{AFH:2002}, we outlined a
programme to systematically study
the six-dimensional $(2, 0)$ theories, with the ultimate goal of eventually
finding an intrinsic definition of them that does not rely on their embedding
into string or $M$-theory.  The $(2, 0)$ theories obey an ADE-classification, 
but have no other discrete or continuous
parameters~\cite{Witten:1995}.  However, such a theory has a moduli
space of
inequivalent vacuum states, parametrized by the expectation values of a set
of scalar fields. At the origin of the moduli space, the theory is given
by some strongly coupled superconformal quantum theory with a global
$R$-symmetry group $G \simeq SO (5)$. In a certain sense, the most 
natural way to study a $(2, 0)$ theory would
be to work directly around this point in the moduli space.  
However, although some intriguing properties of these superconformal
theories are known, it seems fair to say that they are still rather
mysterious and, with present techniques, not easily studied.  

Our approach is instead based on the idea of working at a generic point 
in the moduli space, where the $R$-symmetry is spontaneously broken to a 
subgroup $H$ isomorphic to $SO (4)$. Here, a $(2, 0)$ theory describes
massless particles interacting with tensile strings.
The particles can be obtained by quantizing a set of six-dimensional fields 
known as a $(2, 0)$ tensor multiplet. The tensor multiplet comprises
the scalar fields, whose expectation values parametrize the moduli
space, and their superpartners. The distance to the origin of the
moduli space determines
the tension of the strings, whereas the remaining scalar fields can be
interpreted as the Goldstone bosons associated with the spontaneous breaking
of the $R$-symmetry group from $G \simeq SO (5)$ to $H \simeq SO (4)$. 
(For simplicity, we limit ourselves
in this discussion to the $A_1$ version of $(2, 0)$ theory, which only 
contains a single tensor multiplet and a single type of string.) 
On the world-sheet of a string,
we have a set of two-dimensional fields consisting of the Goldstone
bosons and fermions associated with the further spontaneous breaking of 
the $(2, 0)$ superalgebra by the string.

The dynamics of the theory can be investigated by defining a dimensionless
parameter as the typical energy scale of some process, \eg the energy
of an incoming tensor multiplet quantum, divided by the square root of
the string tension.
In the limit that the string tension goes to infinity, 
so that this parameter goes to zero, the
six-dimensional tensor multiplet fields propagating in
space-time and the two-dimensional fields propagating on the
world-sheets of the strings decouple from each other. The ensuing
free theory is exactly solvable and can serve as a 
starting point for a perturbative analysis of the interacting theory that 
is obtained at a finite string tension.

In this paper, we continue our work along these lines. In the
next section, we review the $(2, 0)$ superalgebra and
its representations on particles and strings. In section three, we
give a Lagrangian for a free tensor multiplet, which to
our knowledge has not yet appeared in the literature. In section four,
we rewrite this field theory in a way appropriate for working
at a generic point in the moduli space with spontaneously broken
$R$-symmetry.
Finally, in section five, we construct the theory of a free string.
In a forthcoming publication, we intend to pursue the challenging problem
of constructing the supersymmetric theory of tensor multiplet particles
interacting with tensile, self-dual strings. We hope that the
results of the present paper will provide a firm basis for this project.

\section{The $(2,0)$ superalgebra and its representations}
The $(2, 0)$ superalgebra~\cite{Nahm:1978} 
contains the six-dimensional Lorentz algebra
$so (5, 1)$ as a subalgebra. The basic representations of the latter algebra,
from which all other representations can be constructed,
are the Weyl spinor ${\bf 4}$ and the anti-Weyl spinor
${\bf 4}^\prime$. We denote quantities transforming in these representations
with a subscript or superscript index $\alpha = 1, \ldots, 4$ respectively.
 The tensor products of two of these representations are
\beqa 
{\bf 4} \otimes {\bf 4} & \simeq & {\bf 6} \oplus {\bf 10}_+ \cr 
{\bf4}^\prime \otimes {\bf 4}^\prime & \simeq & {\bf 6} \oplus {\bf 10}_-
\cr 
{\bf 4} \otimes {\bf 4}^\prime & \simeq & {\bf 1} \oplus {\bf 15} .  
\eeqa 
Here, ${\bf 1}$ is of course the singlet, associated with the invariant
tensor $\de_\al^{\ph{\al}\be}$. The ${\bf 6}$ is the vector,
${\bf 10}_+$ and ${\bf 10}_-$ are the self-dual and anti self-dual
three-forms respectively, and ${\bf 15}$ is the two-form
representation of $so (5,1)$. Finally, we note that the completely
anti-symmetric tensor product of four Weyl spinors or of four
anti-Weyl spinors is a singlet, and we denote the corresponding invariant
tensors as $\epsilon_{\alpha \beta \gamma \delta}$ and 
$\epsilon^{\alpha \beta \gamma \delta}$ respectively. 

The $(2, 0)$ superalgebra also has an $R$-symmetry subalgebra isomorphic to 
$so (5)$. For our purposes, it is convenient to think of $so (5)$ as a
subalgebra of $so (6)$, and use the fact that $so (6) \simeq su (4)$. 
In terms of Lie groups, $SO (6) \simeq SU (4) / {\mathbb Z}_2$. An element 
$g \in SU (4)$ is a four-by-four complex matrix such that $\det g = 1$ and
$g g^\dagger = {\bf 1}_4$. In our case, the most important 
representations of $SU( 4)$, apart from the singlet ${\bf 1}$, are the
${\bf 4}$, the ${\bf 6}$, and the ${\bf 10}$. They correspond to a four-by-one
column vector $\Psi$, an anti-symmetric four-by-four matrix $A$, 
and a symmetric four-by-four matrix $S$ respectively. An element
$g \in SU (4)$ acts on these as
\beqa
\Psi & \rightarrow & \Psi' = g \Psi \\
A & \rightarrow & A' = g A g^{\sss{T}} \\ 
S & \rightarrow & S' = g S g^{\sss{T}} .
\eeqa

Now choose a fixed element $\Om$ in the ${\bf 6}$ representation of $SU (4)$,
\ie an anti-symmetric four-by-four matrix, and normalize it so that
$- \frac{1}{4} {\rm Tr} ( \Om \Om^*) = 1$. Explicitly, we take
\beq
\eqnlab{om_def}
\Om = \pmatrix{ \si_2 & {\bf 0} \cr {\bf 0} & \si_2 \cr } =
\pmatrix{0 & -i & 0 & 0 \cr i & 0 & 0 & 0 \cr 0 & 0 & 0 & -i \cr 0 & 0
  & i & 0 \cr}.
\eeq
We can then define a subgroup
$G \simeq SO(5)$ in $SU(4)$ by the requirement that $g \in G$ if and only if
$g \Om g^{\sss{T}} = \Om$. Hence $\Om$ itself is in the singlet
representation of $G$. We also get a five-dimensional
representation of $G$ on the set of anti-symmetric four-by-four matrices
$\Phi$ obeying $\Tr{\Om \Phi^*} = 0$, so the ${\bf 6}$ representation of 
$SU (4)$ decomposes into the irreps ${\bf 1}$ and ${\bf 5}$ of 
the subgroup $G$. The ${\bf 4}$ representation of $SU (4)$ is still 
irreducible under $G$,  and the tensor product of two such representations 
is given by
\beq
{\bf 4} \otimes {\bf 4} \simeq {\bf 1} \oplus {\bf 5} \oplus {\bf 10} .
\eeq

We are now ready to discuss the complete $(2, 0)$ superalgebra. Its fermionic 
generators $Q_\alpha$ transform in the Weyl representation ${\bf 4}$ under
the Lorentz group and in the ${\bf 4}$ representation under the 
$G \simeq SO (5)$ $R$-symmetry group. They also obey the symplectic Majorana 
reality condition~\cite{Kugo:1983}
\beq
\eqnlab{MW}
Q_\alpha^* = C_{\alh}^{\ph{\alh} \beh} \Om Q_{\beh} .
\eeq
In this relation, the charge conjugation operator $C_\alh^{\ph{\alh} \beh}$ 
is an $SO(5)$ scalar obeying
\beq
{C^*}_{\alh}^{\ph{\alh} \beh} C_{\beh}^{\ph{\beh} \gah} = -
\de_{\alh}^{\ph{\alh}\gah},
\eeq
while $\Om$ is the $SO(5)$ invariant matrix introduced above. Later, we will
also use quantities, such as the parameters $\eta^\alpha$ of infinitesimal 
supersymmetry transformations, that transform in the anti-Weyl representation 
${\bf 4}'$ under the Lorentz group and in the ${\bf 4}$ representation under
$R$-symmetry. Here we introduce the similar reality condition
\beq
\eqnlab{MW2}
\etas{\alh} = - {C^*}_{\beh}^{\ph{\beh} \alh} \Om \eta^{\beh}.
\eeq

According to the above, the anti-commutator of two fermionic generators must
be of the form~\cite{Howe:1998}
\beq
\{ Q_\alpha , Q_\beta^{\sss{T}} \} = P_{\alpha \beta} \Om + Z_{\alpha \beta} 
+ W_{\alpha \beta} .
\eeq
Here, $P_{\alpha \beta} = P_{[\alpha \beta]}$ are scalars under 
$G \simeq SO (5)$, and are interpreted as the six-dimensional momentum 
operators. The generators $Z_{\alpha \beta} = Z_{[\alpha \beta]}$ transform 
in the ${\bf 5}$ representation of $G$, \ie 
$Z_{\alpha \beta} = - Z_{\alpha \beta}^{\sss{T}}$ and 
${\rm Tr} (\Om Z^*_{\alpha \beta}) = 0$. They are `central charges' that
appear in the presence of strings. Finally, the generators
$W_{\alpha \beta} = W_{(\alpha \beta)}$ form another set of central charges
(associated with self-dual three-brane solitons) 
that transform in the ${\bf 10}$ representation of $G$,
\ie $W_{\alpha \beta} = W_{\alpha \beta}^{\sss{T}}$. 
The bosonic subalgebra of the
$(2, 0)$ superalgebra is spanned by $P_{\alpha \beta}$, $Z_{\alpha \beta}$
and $W_{\alpha \beta}$ together with the Lorentz generators and the 
$R$-symmetry generators.

When constructing an irreducible representation of the $(2, 0)$
superalgebra, we may impose any $SO (5, 1) \times SO (5)$ covariant
conditions
on the eigenvalues of the simultaneously diagonalizable generators
$P_{\alpha \beta}$, $Z_{\alpha \beta}$ and $W_{\alpha \beta}$. For example,
we may consider the case when the momentum $P_{\alpha \beta}$ is restricted 
to be light-like and the central charges $Z_{\alpha \beta}$ and
$W_{\alpha \beta}$ vanish. Such a configuration spontaneously breaks the
Lorentz group to the little subgroup $SO (4) \simeq SU (2) \times SU
(2)$, but leaves the $R$-symmetry group $SO(5)$ unbroken.
The supersymmetry generators transform under $SO (4) \times SO (5)
\simeq SU (2) \times SU (2) \times SO (5)$ as
\beq 
({\bf 1}, {\bf 2};{\bf 4}) \oplus ({\bf 2}, {\bf 1};{\bf 4}).
\eeq
It can be seen from the 
supersymmetry algebra that the supercharges of the first term are unbroken
and annihilate a quantum state of this configuration. The
supercharges of the second term form a
Clifford algebra, which can be represented on a set of states transforming as
\beq
({\bf 3}, {\bf 1}; {\bf 1}) \oplus ({\bf 2}, {\bf 1}; {\bf 4})
\oplus ({\bf 1}, {\bf 1}; {\bf 5}).
\eeq
These $8 + 8$ states are known as a $(2, 0)$ tensor particle multiplet, 
because of the states $({\bf 3}, {\bf 1}; {\bf 1})$, that transform
in the rank two antisymmetric self-dual tensor representation of the
unbroken $SO (4)$ subgroup of the Lorentz group.

Another set of conditions can be described as
follows~\cite{Gustavsson:2001sr}: The central charge
$Z_{\alpha \beta}$ is a tensor product of a space-like 
Lorentz vector $V_{\alpha \beta}$ invariant under $R$-symmetry, and a Lorentz 
scalar $\Phi$ in the ${\bf 5}$ representation of the $R$-symmetry group.
The central charge $W_{\alpha \beta}$ vanishes. The scalar product of
$P_{\alpha \beta}$ and $Z_{\alpha \beta}$ is also zero. These conditions
describe a configuration with a straight infinite string pointing in the
direction of $V_{\alpha \beta}$ in a background with the moduli given by
$\Phi$. Unitarity of the representation imposes
that $P_{\alpha \beta}$ be time-like with a certain lower bound on the 
energy (per unit length of the string). We take $P_{\alpha \beta}$ to
saturate this bound. A non-zero $\Phi$ breaks
the $R$-symmetry group spontaneously to $SO (4) \simeq SU (2) \times SU (2)$,
and the vectors $P_{\alpha \beta}$ and $V_{\alpha \beta}$ break the
Lorentz group to $SO (4) \simeq SU (2) \times SU (2)$.
The supersymmetry generators transform as
\beq
({\bf 2}, {\bf 1}; {\bf 2}, {\bf 1}) \oplus 
({\bf 1}, {\bf 2}; {\bf 1}, {\bf 2}) \oplus
({\bf 2}, {\bf 1}; {\bf 1}, {\bf 2}) \oplus
({\bf 1}, {\bf 2}; {\bf 2}, {\bf 1}) ,
\eeq
under $SO (4) \times SO (4) \simeq SU (2) \times SU (2) \times SU(2) 
\times SU (2)$,
where the first $SO (4)$ factor comes from the Lorentz group and the second
from the $R$-symmetry group. 
It can be seen from the supersymmetry algebra that the supercharges of the
last two terms 
are unbroken and annihilate such a state. The supercharges of the first two 
terms form a Clifford algebra, which can be represented on a set of states 
transforming as
\beq
({\bf 2}, {\bf 2}; {\bf 1}, {\bf 1}) \oplus 
({\bf 2}, {\bf 1}; {\bf 1}, {\bf 2}) \oplus
({\bf 1}, {\bf 2}; {\bf 2}, {\bf 1}) \oplus
({\bf 1}, {\bf 1}; {\bf 2}, {\bf 2}) .
\eeq
We call this set of $8 + 8$ states a $(2, 0)$ vector string multiplet,
because of the states $({\bf 2}, {\bf 2}; {\bf 1}, {\bf 1})$
that transform as a vector under the unbroken $SO (4)$ subgroup of the
Lorentz group.

\section{The tensor multiplet field theory}
\subsection{The fields}
\seclab{fields}
The Fock space of the tensor particle multiplet described in the previous
section may be obtained by quantizing a free field theory. We will start
by defining the field content of this theory.

First, we have a set of fermionic fields which are both Lorentz and
$R$-symmetry spinors, \ie transform in the ${\bf 4}$ representation of
$SO(5,1)$ and in the ${\bf 4}$ representation of $SO(5)$. These are
written as (Grassmann odd) column vectors $\Psi_{\alh}$ with four
complex components and obey the symplectic Majorana condition as
defined in \Eqnref{MW}.

Next, we have a set of bosonic fields in the ${\bf 5}$ (vector)
representation of $SO(5)$, transforming as Lorentz scalars. These
fields are represented by the four-by-four antisymmetric matrices
$\Phi$ described above (obeying $\Tr{\Om \Phi^*}=0$) and are subject
to the reality condition
\beq
\Phi^* = - \Om \Phi \Om.
\eeq
The vacuum expectation values of these fields constitute the moduli of
the theory.

Finally, we have a two-form gauge field in the ${\bf 15}$
  representation of $SO(5,1)$, \ie a field $\Bf{\alh}{\beh}$ with
$\Bf{\alh}{\alh} = 0$, which is an $R$-symmetry scalar. Infinitesimal
  gauge transformations with a one-form parameter $\La^{\alh
\beh} = - \La^{\beh \alh}$ act as
\beq
\de \Bf{\alh}{\beh} = \pa_{\alh\gah} \La^{\gah\beh} -
\frac{1}{4}\kde{\alh}{\beh} \pa_{\deh\gah} \La^{\gah\deh},
\eeq
where $\pa_{\alh \beh} = \pa_{[\alh \beh]}$ are the space-time
derivatives. For convenience, we also define a derivative with
  superscript
indices through $\pa^{\alh \beh} \equiv \frac{1}{2} \eps^{\alh \beh
\gah \deh} \pa_{\gah \deh} $. The gauge transformations leave the
following field strength components invariant:
\beqa
H_{\alh \beh} & = & \pa_{\alh \gah} \Bf{\beh}{\gah} + \pa_{\beh \gah}
\Bf{\alh}{\gah} \\
H^{\alh \beh} & = & \pa^{\alh \gah} \Bf{\gah}{\beh} +
\pa^{\beh \gah} \Bf{\gah}{\alh},
\eeqa
corresponding to the self-dual and the anti self-dual three-form
representations of $SO(5,1)$ respectively. Actually, the tensor
multiplet only contains the self-dual part $H_{\alh \beh}$ of the
field strength. However, it is not possible to write down an action
for a self-dual three-form~\cite{Witten:1997}, so in order to give a
Lagrangian description of it, we relax the self-duality requirement by
including also the anti self-dual part $H^{\alh \beh}$ as a ``spectator
field'' in the theory. This means that $H^{\alh \beh}$ in no way mixes
  with the other fields of the tensor multiplet. When including string
  interactions, the coefficients of the coupling terms in the action
  are to be chosen such that this field does not couple.

The derivatives of the field strength
components are related by the Bianchi identity 
\beq
\pa^{\alh \gah} H_{\alh \beh} - \pa_{\alh \beh} H^{\alh \gah} =
0,
\eeq
which is easily verified by insertion.

The gauge field and the field strengths obey the reality conditions
\beqa
{B^*}_{\alh}^{\ph{\alh}\beh} & = & C_{\alh}^{\ph{\alh} \ga} (-
{C^*}_{\de}^{\ph{\de} \beh}) B_{\ga}^{\ph{\ga}\de} \\
H^*_{\alh \beh} & = & C_{\alh}^{\ph{\alh} \ga} C_{\beh}^{\ph{\beh}
  \de} H_{\ga \de} \\
{H^*}^{\alh \beh} & = & (- {C^*}_{\ga}^{\ph{\ga} \alh}) (-
{C^*}_{\de}^{\ph{\de} \beh}) H^{\ga \de}
\eeqa
in analogy with the conditions \eqnref{MW} and \eqnref{MW2} above.

We should also note how the fields of the tensor multiplet transform
under the $R$-symmetry group $G \simeq SO(5)$. For $g \in G$, the
fields transform as
\beqa
\Psi_{\alh} & \rightarrow & g \Psi_{\alh} \\
\Phi & \rightarrow & g \Phi g^{\sss{T}} \\
\Bf{\alh}{\beh} & \rightarrow & \Bf{\alh}{\beh}.
\eeqa
Since $\Om$ is an $SO(5)$-invariant matrix, we also have the
transformation
\beq
\Om \rightarrow g \Om g^{\sss{T}} = \Om.
\eeq

The supersymmetric and $SO(5)$ invariant action governing the
dynamics of the free $(2,0)$ tensor multiplet is
\beq
\eqnlab{(2,0)action}
S = \int d^6x \left\{ \Tr{\pa_{\alh \beh} \Phi \pa^{\alh \beh} \Phi^{*}} +
  2 H_{\alh \beh} H^{\alh \beh} - 4i \PsiT{\alh} \Om \pa^{\alh
    \beh} \Psi_{\beh} \right\},
\eeq
where we, as noted above, are forced to include the anti self-dual part of $H$.
Under arbitrary variations of the fields $B_{\alh}{}^{\beh}$,
$\Phi$ and $\Psi_{\alh}$, the action yields the equations of motion
\beqa
\pa^{\alh \beh} \pa_{\alh \beh} \Phi & = & 0  \eqnlab{phi_eom}\\
\pa^{\alh \gah} H_{\alh \beh} + \pa_{\alh \beh} H^{\alh \gah} & = & 0
\eqnlab{H_eom} \\
\pa^{\alh \beh} \Psi_{\beh} & = & 0 \eqnlab{psi_eom}.
\eeqa
In~\cite{Howe:1983}, these are obtained by a superspace approach at
the level of equations of motion.

\subsection{The supersymmetry transformations}
The supersymmetry transformations of the fields in the tensor
multiplet are
\beqa
\eqnlab{susy_phi}
\de \Phi & = & i \eta^{\alh} \PsiT{\alh} + i \Psi_{\alh}
\etaT{\alh} + \frac{i}{2} (\etaT{\alh} \Om
\Psi_{\alh}) \Om \\
\eqnlab{susy_b}
\de \Bf{\alh}{\beh} & = & i \etaT{\beh} \Om \Psi_{\alh} -
\frac{i}{4} \de_{\alh}^{\ph{\alh}\beh} \etaT{\gah} \Om
\Psi_{\gah} \\
\eqnlab{susy_psi}
\de \Psi_{\alh} & = & H_{\alh \beh} \eta^{\beh} + 2 \pa_{\alh \beh}
\Phi \Om \eta^{\beh},
\eeqa
where $\eta^{\alh}$ is a constant fermionic parameter. The variations
of the fields obey the same reality and $SO(5)$ properties as the fields
themselves. The resulting transformation of the self-dual field
strength becomes
\beq
\eqnlab{susy_sdH}
\de H_{\alh \beh} = i \etaT{\gah} \Om (\pa_{\alh \gah}
\Psi_{\beh} + \pa_{\beh \gah} \Psi_{\alh}).
\eeq
It is also worthwhile to compute the transformation of the anti
self-dual part of the field strength
\beq
\eqnlab{susy_asdH}
\de H^{\alh \beh} = i \etaT{\alh} \Om \pa^{\beh \gah}
\Psi_{\gah} + i\etaT{\beh} \Om \pa^{\alh \gah} \Psi_{\gah}.
\eeq
It follows from \Eqnref{psi_eom} that this variation vanishes
when imposing the equations of motion for $\Psi$. We also note that
$H^{\alh \beh}$ does not appear in the right-hand sides of
\Eqsref{susy_phi} -- \eqnref{susy_psi}. This shows that
$H^{\alh \beh}$ is not really part of the tensor multiplet.

One may verify that the supersymmetry algebra closes
on-shell, \ie the commutator of two supersymmetry transformations
yields a translation according to
\beq
[\de_1,\de_2] = 2i \etaTe{\alh} \Om \eta_2^{\beh}
\pa_{\alh \beh},
\eeq
when acting on any of the fields in the tensor multiplet. Two
supersymmetry transformations acting on the anti self-dual part
$H^{\al \be}$ of the field strength is zero on-shell.

If we perform a \emph{local} supersymmetry variation of the action
\eqnref{(2,0)action} we obtain
\beq
\de S = \int d^6 x \left(i \pa^{\alh \beh} \eta^{\gah}(x)
\right)^{\sss{T}} \Om \left\{8 H_{\gah \alh} \Psi_{\beh} - 16
  \pa_{\gah \alh} \Phi \Om \Psi_{\beh}
\right\}.
\eeq
This means that the Noether currents $(Q_{\gah})_{\alh \beh}$
corresponding to the supersymmetries are given by
\beq
(Q_{\gah})_{\alh \beh} = 8 H_{\gah [ \alh} \Psi_{\beh ]} - 16 \pa_{\gah
  [ \alh} \Phi \Om \Psi_{\beh]}
\eqnlab{freetensorcurrent}
\eeq
and are conserved, \ie $\pa^{\alh \beh} (Q_{\gah})_{\alh \beh} = 0$
when the equations of motion are imposed together with the Bianchi
identity.

\section{Spontaneous symmetry breaking}
\seclab{breaking}
We next consider a configuration where the $R$-symmetry group is
spontaneously broken from $SO(5)$ to $SO(4)$ by the expectation values
of the scalar fields $\Phi$, which are the moduli of the model. This
breaks four continuous symmetries and therefore gives rise to
four Goldstone bosons. We will mainly follow the procedure outlined in
Ch.~19 of~\cite{Weinberg_2}.

Since $\Phi$ has five linearly independent components, we may think of
it as a vector in a five-dimensional space. An $SO(5)$ transformation
rotates this vector. By giving the components of $\Phi$
 non-zero vacuum expectation values we choose a direction
in this space, which breaks the $SO(5)$ symmetry spontaneously to
$SO(4)$ (rotations in the transverse space). This means that the
$SO(5)$ symmetry will be realized non-linearly, and the coset space
$SO(5)/SO(4)$ is parametrized by the Goldstone bosons.

\subsection{The action}
Mathematically, the symmetry breaking is imposed by choosing a fixed
matrix $\Phih$ in the ${\bf 5}$ representation of $G \simeq SO(5)$
such that $ -\frac{1}{4} {\rm Tr} (\Phih \Phih^*) = 1$, and 
defining a subgroup $H \simeq SO(4)$ in $G$ by requiring that $h \Phih
h^{\sss{T}} = \Phih$ for all $h \in H$. This corresponds to choosing a
fixed ``direction'' for the expectation values of the $\Phi$
field. Since $H \simeq SO(4) \simeq SU(2) \times SU(2)$, all matrices
in $H$ are block-diagonal with two $SU(2)$ matrices as diagonal
elements. Explicitly, we choose our coordinates such that
\beq
\Phih = \pmatrix{ \si_2 & {\bf 0} \cr {\bf 0} & -\si_2 \cr } =
\pmatrix{0 & -i & 0 & 0 \cr i & 0 & 0 & 0 \cr 0 & 0 & 0 & i \cr 0 & 0
  & -i & 0 \cr}
\eeq
in the basis where the $SO(5)$ invariant matrix $\Om$ is given by
\Eqnref{om_def}.

Now define the right coset space $G/H$ by identifying the group
element $g\in G$ with $gh$ for $h\in H$. Let $\ga=\ga(\xi)$ be a
representative element in this
four-dimensional space, which is parametrized by the Goldstone boson
fields $\xi$. Any $\Phi$ in the ${\bf 5}$ irrep of $G$ may now be
written as
\beq
\eqnlab{phitransf}
\Phi = \phi \ga \Phih \ga^{\sss{T}},
\eeq
where the real valued field $\phi$ is given by 
$\phi^2 = -\frac{1}{4} {\rm Tr} (\Phi \Phi^*)$.
(Note that
$\ga \ga^{\dagger} = \ga^{\dagger} \ga = {\bf 1}_4$, \ie $\ga$ is a
unitary matrix.) Hence, the vacuum expectation value $\phivev$ is a
measure of how far we are from the origin of the moduli
space and therefore sets the scale of the symmetry
breaking. Analogously, we define a new spinor field $\Psit$ by
\beq
\eqnlab{psitransf}
\Psi = \ga \Psit.
\eeq
This field transforms linearly under $H \simeq SO(4)$, \ie $\Psit
\rightarrow h \Psit$ for $h\in H$, and non-linearly under the full
$G\simeq SO(5)$ group.

We may now insert the expressions \eqnref{phitransf} and
\eqnref{psitransf} in the action \eqnref{(2,0)action}. Starting with
the kinetic term for the fermionic field, we get
\beq
-4i \PsiT{\alh} \Om \pa^{\alh \beh} \Psi_{\beh} = -4
\PsitT{\alh} \Om \left( i\pa^{\alh \beh} + i \ga^{-1}
  \pa^{\alh \beh} \ga \right) \Psit_{\beh}.
\eeq
The Lie algebra element $i \ga^{-1} \pa^{\alh \beh} \ga$ may be
decomposed according to
\beq
\eqnlab{a_pi}
i \ga^{-1} \pa^{\alh \beh} \ga = A^{\alh \beh} + \Pi^{\alh \beh},
\eeq
where $A^{\alh \beh}$ consists of the part belonging to the Lie
algebra of $H$, and $\Pi^{\alh \beh}$ belongs to its orthogonal
complement in the Lie algebra of $G$. In other words, $A^{\alh \beh}$
is the block-diagonal part
and $\Pi^{\alh \beh}$ is off-block-diagonal. Actually, $\Pi^{\alh
  \beh}$ is the covariant derivative of the Goldstone fields. It is
useful to note that $A^{\alh \beh}$ commutes with the matrix $\Phih
\Om=\Om \Phih$, while $\Pi^{\alh \beh}$ anti-commutes.

Using this expansion, we may introduce a covariant derivative $D^{\alh
  \beh}$ which acts on an $SO(4)$ covariant spinor according to
\beq
D^{\alh \beh} \Psit_{\beh} = \left( \pa^{\alh \beh} - i A^{\alh \beh}
\right) \Psit_{\beh};
\eeq
this quantity transforms in the same way as $\Psit$ under a $G$
transformation. 

Expanding the kinetic term for the $\Phi$ field in the same way, the
action in the new set of variables becomes
\beqa
S & = & \int d^6x \Big\{ -4 \pa_{\alh \beh} \phi \pa^{\alh \beh} \phi - 4 \phi^2
  \Tr{\Pi_{\alh \beh} \Pi^{\alh \beh}} + 2 H_{\alh \beh} H^{\alh
    \beh} - {} \nn \\ 
\eqnlab{brokenaction}
& & {} - 4 i \PsitT{\alh} \Om D^{\alh \beh} \Psit_{\beh} - 4
\PsitT{\alh} \Om \Pi^{\alh \beh} \Psit_{\beh}  \Big\}.
\eeqa
This is the form of the action that we intend to use when coupling the
tensor multiplet to a string. The action is manifestly $SO(4)$
invariant, but $SO(5)$ invariance in ensured by the use of covariant
derivatives. 

By varying this action with respect to $B$, $\phi$, $\Psit$ and the
Goldstone boson fields $\xi$, we obtain the equations of motion
\beqa
\pa_{\alh \beh} \pa^{\alh \beh} \phi - \phi \Tr{\Pi_{\alh \beh}
  \Pi^{\alh \beh}} & = & 0 \\
\phi D_{\alh \beh} \Pi^{\alh \beh} + 2 \pa_{\alh \beh} \phi \Pi^{\alh \beh}
& = & 0 \\
\pa^{\alh \gah} H_{\alh \beh} + \pa_{\alh \beh} H^{\alh \gah} & = & 0
\\
(iD^{\alh \beh} + \Pi^{\alh \beh}) \Psit_{\beh} & = & 0.
\eeqa
The equations of motion can also be obtained by a change of
variables in \Eqsref{phi_eom} -- \eqnref{psi_eom}. Note that the first
two equations above both follow from \Eqnref{phi_eom}.

\subsection{The supersymmetry transformations}
The next step is to derive the supersymmetry transformations of $H$, $\phi$,
$\Pi$ and $\Psit$. For this purpose, we define $\etat$ through
\beq
\eta = \ga \etat.
\eeq
The transformation of $H$ follows from a simple
rewriting of \Eqnref{susy_sdH} and becomes
\beq
\de H_{\alh \beh} = \etatT{\gah} \Om \left( (i D_{\alh
    \gah}+\Pi_{\alh \gah} ) \Psit_{\beh} + (i D_{\beh \gah}+\Pi_{\beh
    \gah}) \Psit_{\alh} \right),
\eeq
while the anti self-dual part transforms as
\beq
\de H^{\alh \beh} = \etatT{\alh} \Om (i D^{\beh \gah} +
\Pi^{\beh \gah} ) \Psit_{\gah} + \etatT{\beh} \Om (i D^{\alh
  \gah}+\Pi^{\alh \gah}) \Psit_{\gah}.
\eeq
The transformation of $\phi$ follows from the equation 
$\phi^2 = -\frac{1}{4} {\rm Tr} (\Phi \Phi^*)$ 
combined with \Eqnref{susy_phi}. The result is
\beq
\de \phi = - \frac{i}{2} \etatT{\alh} \Phih \Psit_{\alh}.
\eeq
To proceed, we need to consider the supersymmetry
transformation of $i\ga^{-1} \pa^{\alh \beh} \ga$. This obeys the
identity
\beq
\eqnlab{depa}
\de (i\ga^{-1} \pa^{\alh \beh} \ga) = i \com{ (i\ga^{-1} \de
  \ga)}{(i\ga^{-1} \pa^{\alh \beh} \ga)} + \pa^{\alh \beh} (i\ga^{-1}
\de \ga).
\eeq
In analogy with the decomposition \eqnref{a_pi} we write
\beq
i\ga^{-1} \de \ga = S + T,
\eeq
where $S$ denotes the block-diagonal part and $T$ the
off-block-diagonal part. By transforming both sides of
\Eqnref{phitransf} and comparing with \Eqnref{susy_phi}, we find that
\beq
T = - \frac{1}{4\phi} \left( (\etat^{\gah} \PsitT{\gah} +
  \Psit_{\gah} \etatT{\gah} ) \Phih - \Om \Phih (\etat^{\gah}
  \PsitT{\gah} + \Psit_{\gah} \etatT{\gah} ) \Om
\right).
\eeq
An explicit expression for $S$ is never needed in the calculations.

Hence, using \Eqnref{a_pi}, we may decompose \Eqnref{depa} as
\beqa
\de A^{\alh \beh} & = & i \com{S}{A^{\alh \beh}} + i \com{T}{\Pi^{\alh
    \beh}} + \pa^{\alh \beh} S \\
\de \Pi^{\alh \beh} & = & i \com{T}{A^{\alh \beh}} + i
\com{S}{\Pi^{\alh \beh}} + \pa^{\alh \beh} T.
\eeqa
It is convenient to introduce a covariant supersymmetry variation
$\De$ acting as
\beqa
\De \Psit_{\alh} & = & \de \Psit_{\alh} - i S \Psit_{\alh} \\
\De \Pi^{\alh \beh} & = & \de \Pi^{\alh \beh} - i \com{S}{\Pi^{\alh
    \beh}}.
\eeqa
This yields that
\beq
\De \Pi^{\alh \beh} = \pa^{\alh \beh} T - i \com{A^{\alh \beh}}{T}
\equiv D^{\alh \beh} T,
\eeq
where we have introduced the covariant derivative acting on $T$. We
also find that
\beq
\De \Psit_{\alh} = i T \Psit_{\alh} + H_{\alh \beh} \etat^{\beh} + 2
\pa_{\alh \beh} \phi \Phih \Om \etat^{\beh} + 4i\phi \Phih \Om
\Pi_{\alh \beh} \etat^{\beh}
\eeq
and
\beq
\De (D^{\alh \beh} \Psit_{\gah}) = D^{\alh \beh} (\De \Psit_{\gah} ) +
\com{T}{\Pi^{\alh \beh}} \Psit_{\gah}.
\eeq

Having found the supersymmetry transformations of the fields, we now
consider the action \eqnref{brokenaction}. It is straightforward to
verify that
\beqa
\de \Tr{\Pi_{\alh \beh} \Pi^{\alh \beh}} & = & 2 \Tr{\De \Pi_{\alh \beh}
  \Pi^{\alh \beh}} \\
\de (\PsitT{\alh} \Om D^{\alh \beh} \Psit_{\beh}) & = &
(\De\Psit_{\alh})^{\sss{T}} \Om D^{\alh \beh} \Psit_{\beh} +
\PsitT{\alh} \Om \De(D^{\alh \beh} \Psit_{\beh}) \\
\de (\PsitT{\alh} \Om \Pi^{\alh \beh} \Psit_{\beh}) & = &
2 \PsitT{\alh} \Om \Pi^{\alh \beh} \De \Psit_{\beh} +
\PsitT{\alh} \Om (\De \Pi^{\alh \beh}) \Psit_{\beh},
\eeqa
by using that $\Tr{\com{S}{\Pi} \Pi} =0$ and that $S^{\sss{T}} \Om +
\Om S = 0$. It follows that the action \eqnref{brokenaction} is, as
expected, invariant under the supersymmetry transformations. In this
calculation, it is necessary to use that
\beqa
D_{\alh \beh} \Pi^{\alh \gah} & = & \frac{1}{4} \de_{\beh}^{\ph{\beh}
  \gah} D_{\alh \deh} \Pi^{\alh \deh} \\
\Pi_{\alh \beh} \Pi^{\alh \beh} & = & \frac{1}{4} \Tr{\Pi_{\alh \beh}
  \Pi^{\alh \beh}} {\bf 1}_4.
\eeqa
We also apply the fact that the supersymmetry transformation is
global, \ie $\pa \eta = 0$, which gives that
\beq
\eqnlab{D_etat}
D_{\alh \beh} \etat^{\beh} = i \Pi_{\alh \beh} \etat^{\beh}.
\eeq

For future reference, we note that a local supersymmetry variation of
the action yields the Noether currents
\beq
(\tilde{Q}_{\gah})_{\alh \beh} = 8 H_{\gah [ \alh} \Psit_{\beh ]} -
32i\phi \Phih \Om \Pi_{\gah [ \alh} \Psit_{\beh ]} - 16 \pa_{\gah
  [ \alh} \phi \Phih \Om \Psit_{\beh]},
\eeq
obeying the conservation equation
\beq
(iD^{\alh \beh} + \Pi^{\alh \beh}) (\tilde{Q}_{\gah})_{\alh \beh} = 0.
\eeq

\section{The free string}
\subsection{The fields}
A configuration containing a straight static string breaks translational 
invariance in the directions transverse to the string and also half of 
the supersymmetries.
The corresponding Goldstone modes consist of four bosons transforming
as $({\bf 2}, {\bf 2}; {\bf 1}, {\bf 1})$, four left moving fermions
transforming as $({\bf 2}, {\bf 1}; {\bf 2}, {\bf 1})$, and four right
moving fermions transforming as $({\bf 1}, {\bf 2}; {\bf 1}, {\bf 2})$
under $SO (4) \times SO (4) \simeq 
SU (2) \times SU (2) \times SU (2) \times SU (2)$. Here, the first and 
second $SO (4)$ factors are the unbroken parts of the Lorentz and
$R$-symmetry groups respectively. Quantization of the zero-modes of
these fields gives rise to the vector string multiplet described in
section two. Quantization of their non zero-modes gives rise to a Fock
space of waves propagating on the world-sheet of the string.

To construct such a string theory, it is convenient to supplement these
world-sheet Goldstone fields with extra
components so that they fill out representations of the unbroken
Lorentz group $SO (5, 1)$. We will thus take the fields on the
string world-sheet to be a space-time vector 
$X^{\alpha \beta} = X^{[\alpha \beta]}$ which is an $R$-symmetry singlet,
and a space-time spinor $\thetat^\alpha$ in the
$({\bf 2}, {\bf 1}) \oplus ({\bf 1}, {\bf 2})$ representation of the
unbroken subgroup $SO (4) \simeq SU (2) \times SU (2)$ of the
$R$-symmetry group. The action governing these fields
should be invariant under local reparametrizations of the world-sheet
and a local fermionic `$\kappa$-symmetry' by which the added components
can be eliminated. It is also convenient to introduce an auxiliary
world-sheet metric $g_{i j}$, where $i, j = 1, 2$ are world-sheet
vector indices. This field obeys an algebraic
equation of motion~\cite{Brink:1976,Deser:1976}, and thus does not
contain any independent degrees of freedom.

The action should be invariant under all transformations
of the $(2, 0)$ superalgebra. However, only the unbroken $SO (4)$ subgroup of
the $R$-symmetry group will be linearly realized. The full $SO (5)$ 
$R$-symmetry group can then be non-linearly realized by replacing
all derivatives with covariant derivatives involving the space-time
Goldstone fields discussed in section four. We will not discuss such
couplings between space-time and world-sheet fields in this paper, though,
but defer them to a forthcoming publication. This also applies to the
string tension, which we will simply take to be given by the vacuum
expectation value $\phivev$ of $\phi$, although it should be
given by the local value of $\phi$ in the complete theory.

\subsection{The action and its symmetries}
The construction of the theory is largely analogous to the Green-Schwarz
superstring~\cite{Green:1984}. Global supersymmetry transformations
act as
\beqa
\delta \thetat^\alpha & = & \etat^\alpha \\
\delta X^{\alpha \beta} & = & 
i \etatT{[\alpha} \Omega \thetat^{\beta]} \\
\delta g_{i j} & = & 0,
\eeqa
where we take the parameters $\etat^\alpha$ to be constant. In the complete
theory, they should fulfill a requirement like \Eqnref{D_etat}. We define a
`conjugate momentum' as 
\beq
P_i^{\alpha \beta} = \partial_i X^{\alpha \beta} 
- i \thetatT{[\alpha} \Omega \partial_i \thetat^{\beta]} 
\eeq
and find that
\beq
\delta P_i^{\alpha \beta} = 0 .
\eeq

Denoting the world-sheet coordinates by $\si^i$, the term
\beq
S_1 = - \frac{1}{8} \phivev
\int d^2 \sigma \sqrt{g} g^{i j} \eps_{\al \be \ga \de} P_i^{\al \be}
P_j^{\ga \de},
\eeq
where $g^{i j}$ is the inverse of $g_{i j}$ and $g = - \det g_{i j}$, is
obviously invariant under supersymmetry. We also consider the term
\beq
S_2 = - \frac{1}{4} \phivev \int d^2 \sigma \epsilon^{i j} 
\epsilon_{\alpha \beta \gamma \delta} \left( i P_i^{\alpha \beta}
\thetatT{\gamma} \Phih \partial_j \thetat^\delta - \frac{1}{2}
\thetatT{\alpha} \Om \partial_i \thetat^\beta 
\thetatT{\gamma} \Phih \partial_j \thetat^\delta \right) .
\eeq
Using the constancy of the parameters $\etat^\alpha$ and discarding total
derivatives, it is straightforward to show that
\beq
\eqnlab{S_2}
\delta S_2 = \frac{1}{4} \phivev \int d^2 \sigma \epsilon^{i j}
\epsilon_{\alpha \beta \gamma \delta} \lambda_{[1}^{{\sss T} \alpha} \Omega 
\lambda_2^\beta \lambda_3^{{\sss T} \gamma} \Phih \lambda_{4]}^\delta ,
\eeq
where
$\lambda_1^\alpha = \etat^\alpha$, $\lambda_2^\beta = \thetat^\beta$,
$\lambda_3^\gamma = i \partial_i \thetat^\gamma$,
$\lambda_4^\delta = i \partial_j \thetat^\delta$. But by direct computation,
one can verify the identity
\beq
\eqnlab{Fierz}
v_{[1}^{{\sss T}} \Omega  v_2 v_3^{{\sss T}} \Phih v_{4]}= 0,
\eeq
which is valid for any Grassmann \emph{even} quantities
$v_1,\ldots,v_4$ in the $({\bf 2},{\bf 1}) \oplus ({\bf 1},{\bf 2})$
representation of $SO(4)$. It follows that the integrand of
\Eqnref{S_2}, which involves Grassmann \emph{odd} quantities
$\la_1^{\al}, \ldots,\la_4^{\de}$, vanishes because of the
antisymmetry of $\eps_{\al \be \ga \de}$. So the term $S_2$ is
invariant under supersymmetry.

The local fermionic symmetry has parameters $\tilde{\kappa}^i_\alpha$ 
that are functions of the world-sheet coordinates $\sigma^i$. 
They are subject to the constraint
\beq
\tilde{\kappa}^i_\alpha = {\cal P}^{i j} g_{j k} \tilde{\kappa}^k_\alpha ,
\eeq
where the projection operator ${\cal P}^{i j}$ is given by
\beq
{\cal P}^{i j} = \frac{1}{2} \left( g^{i j} {\bf 1}_4 
- \epsilon^{i j} \Om \Phih / \sqrt{g} \right) .
\eeq
The transformations act on the fields as
\beqa
\delta_{\kappa} \thetat^\alpha & = & P_i^{\alpha \beta}
\tilde{\kappa}^i_\beta \\ 
\delta_{\kappa} X^{\alpha \beta} & = & - i \delta_{\kappa}
\thetatT{[\alpha} \Om  \thetat^{\beta]} \\
\delta_{\kappa} \left( \sqrt{g} g^{i j} \right) & = & 2 i \sqrt{g} 
\left( {\cal P}^{i k} \tilde{\kappa}^j_\alpha \right)^{\sss{T}} 
\Om \partial_k \thetat^\alpha .
\eeqa
The last equation is consistent with the symmetry and unimodularity of
$\sqrt{g} g^{i j}$. It follows from the first two equations that
\beq
\delta_{\kappa} P_i^{\alpha \beta} = - 2 i \delta_{\kappa}
\thetatT{[\alpha} \Om \partial_i \thetat^{\beta]} .
\eeq
The terms $S_1$ and $S_2$ are not separately invariant under these
transformations. However, by using \Eqnref{Fierz},
one can verify that the variation of the linear combination
\beq
S = S_1 + S_2
\eeq
vanishes. This is our free string action. It should however
be coupled to a space-time tensor multiplet so that the string tension is
given by the local value of $\phi$ rather than by $\left< \phi \right>$,
and so that ordinary derivatives are replaced by covariant derivatives
involving the pullback of the space-time Goldstone fields to the world-sheet.
The string should also couple electrically and magnetically to the two-form
gauge field of the tensor multiplet. Constructing such a theory while
maintaining supersymmetry and $\kappa$-symmetry is a challenging problem,
on which we hope to make further progress.

\vspace{1cm}
\noindent
\textbf{Acknowledgments:} M.H.~is a Research Fellow at the Royal
Swedish Academy of Sciences. He would like to thank the Institute for
Advanced Study in Princeton for its hospitality, and Edward Witten for
an illuminating discussion on spontaneous symmetry breaking. We would
also like to thank Martin Cederwall for many helpful discussions on
supersymmetry.

\clearpage
\bibliographystyle{utphysmod3b}
\bibliography{free}

\end{document}